# Origin of Experimental Realization of Vector Beams by Superposition Technique: Geometric Phase

A. Srinivasa Rao[1,2,3,4]

[1]Graduate School of Engineering, Chiba University, Chiba 263-8522, Japan
[2]Molecular Chirality Research Centre, Chiba University, Chiba 263-8522, Japan
[3]Institute for Advanced Academic Research, Chiba University, Chiba, 263-8522, Japan
[4]Quantlight and High Harmonics Lab Pvt. Ltd., Door No. 3-7-400/PGRC/304, Survey No. 30/P, Street No. 16, Nalanda Nagar, Hyderguda, Hyderabad 500048, India
*asvrao@chiba-u.jp*, *sri.jsp7@gmail.com*

**Abstract**

Optical vector beam generation based on the superposition technique [Phys. Rev. Lett. **107**(5), 053601 (2011)] has attracted significant interest in both fundamental and applied optics due to its simplicity and cost-effectiveness. In this approach, the superposed modes exist in orthogonal polarization states and therefore do not interfere with each other. When optical gadgets are employed for characterization, the corresponding polarization distribution can be observed. In this work, we report that the characteristic (non-uniform) polarization distribution of vector beams is not inherently present in their field distribution. Instead, the observed polarization features arise during characterization as a consequence of the geometric phase difference introduced between the superposed modes by the optical gadgets. This analysis opens a new direction in the understanding of various classes of vector beams and offers potential advances in the field of structured beam optics.

The geometric phase of light is an intriguing property that manifests when light undergoes cyclic or noncyclic transformations. The optical geometric phase was first introduced in 1956 by Shivaramakrishnan Pancharatnam while analyzing the interference of optical waves with different polarization states on the Poincaré sphere (PS) [1]. At the time, this phase was not recognized by the scientific community as a geometric phase. Subsequently, in 1984, Sir Michael Berry identified the geometric phase in quantum systems [2]. Following this discovery, Ramaseshan and Nityananda discussed Pancharatnam's work with Berry and demonstrated the equivalence between Pancharatnam's classical geometric phase and Berry's quantum geometric phase [3,4]. Since then, the optical geometric phase of light has commonly been referred to as the Pancharatnam–Berry (PB) phase.

The geometric phase of light has been successfully quantified through experimental measurements by inducing changes in various light properties, such as spin angular momentum (SAM), orbital angular momentum (OAM), combined SAM–OAM states, and propagation direction [5]. Over the past few decades, the PB phase has been exploited in a wide range of applications. For example, it has been used as an optical tool to control the phase of atom interferometers formed via diffraction of atomic waves by laser standing waves in the Bragg regime [6], in the fabrication of spin-selective deflectors and focusing lenses [7,8], and in the development of compact diffractive elements such as $q$-plates based on intrinsic spin–orbit interaction arising from PB phase gradients to generate optical vortices [9,10]. Additional applications include phase shifters for optical circuits [11], PB-phase-controlled optical frequency shifters [12], and others. In this work, we introduce an additional significant application of the PB phase in the generation of optical vector beams.

Optical vector beams [13], which exhibit non-uniform polarization distributions across the beam cross-section, have attracted considerable attention due to their unique applications. These include surface interactions with polymers for producing surface relief structures [14], ultrafast laser-induced periodic surface structures (LIPSS) on metal films [15], controlled optical trapping [16], and high-resolution applications such as super-resolution microscopy and microscale drilling using tightly focused radially polarized vector beams that generate strong longitudinal field components and reduced focal spot sizes [17,18]. The fundamental principle underlying most experimental configurations for the realization of optical vector beams is based on superposition, wherein two orthogonal spatial modes are combined in mutually orthogonal polarization states [19].

In this Letter, we demonstrate that when optical elements are employed for the experimental characterization of vector beams generated using the superposition technique, the observed polarization signatures arise as a consequence of the PB phase induced between the superposed optical vortices. This phase originates from the transport of polarization states on the PS during the characterization process.

All possible polarization states of light form a two-dimensional (2D) Hilbert space. These states can be represented on the surface of a two-sphere, known as the Poincaré sphere (PS) [20]. The PS is constructed using the components of the Stokes vector as Cartesian coordinates, and any polarization state $P$ on this sphere can be uniquely represented by the spherical coordinates $(\theta, \phi)$. We refer to this representation as the polarization PS, as shown in Fig. 1($a$). The generalized polarization state on the polarization PS can be mathematically expressed in terms of orthogonal circular polarization states located at the poles as

$$|P(\theta,\phi)\rangle = \cos\left(\frac{\theta}{2}\right)\cdot e^{-\frac{i\phi}{2}}\cdot f_1(r,\phi,z)|LC\rangle + \sin\left(\frac{\theta}{2}\right)\cdot e^{\frac{i\phi}{2}}\cdot f_2(r,\phi,z)|RC\rangle. \tag{1}$$

The angle $\theta\,(0 \leq \theta \leq \pi)$ is referred to as the polar angle and is measured from the north pole. The second angular coordinate, $\phi\,(0 \leq \phi \leq 2\pi)$, is the azimuthal angle and is defined with respect to the $S_1$–$S_3$ plane. The ket vectors $|LC\rangle$ and $|RC\rangle$ represent the left circularly polarized (LCP) and right circularly polarized (RCP) states, respectively. The wave function $f_i(r,\phi,z)$ describes the spatial distribution of the laser mode. When a polarization state undergoes transport along a closed loop on the polarization PS, it acquires a PB phase $\gamma_P$, which is quantitatively given by $\gamma_P = -s\Omega/2$, where $\Omega$ is the solid angle enclosed by the closed loop on the PS and $s$ denotes the spin angular momentum (SAM) of the light. Now, let us consider the superposition of two laser beams in the LCP and RCP states with equal amplitudes (weight factors), mathematically expressed as

$$|P(\pi/2,\phi)\rangle = \frac{1}{\sqrt{2}}\left\{e^{-\frac{i\phi_d}{2}}\cdot f_1(r,\phi,z)|LC\rangle + e^{\frac{i\phi_d}{2}}\cdot f_2(r,\phi,z)|RC\rangle\right\}. \tag{2}$$

This superposition state contains the poles of the PS and it is a non-interfering pure state. The relative phase difference between the two constituent states is the dynamic phase $\phi_d$, which can be continuously tuned by controlling the optical path difference between the two polarization modes. When this superposition state is passed through a polarizer $\hat{P}(\alpha)$, whose transmission axis is oriented at an angle $\alpha$ with respect to the horizontal direction, the resulting output state becomes

$$\hat{P}(\alpha)|P(\pi/2,\phi)\rangle = \frac{1}{2}\{f_1(r,\phi,z) + f_2(r,\phi,z)\cdot e^{i(\phi_d+2\alpha)}\}|L(\alpha)\rangle. \tag{3}$$

Here, the linear polarizer with its transmission axis oriented at an angle $\alpha$ transforms any polarization state on the PS into a linear polarization state located at the coordinate $(\theta = \pi/2, \phi = 2\alpha)$. The spatial modes contained within the curly brackets are projected onto a single linear polarization state and therefore interfere with each other, producing an interference pattern whose intensity distribution is governed by the phase term $\phi_d + 2\alpha$. For a fixed dynamical phase, the evolution of the fringe pattern depends entirely on the polarizer angle $\alpha$. A full rotation of the polarizer $(2\pi)$ results in a $4\pi$ phase shift. In this context, the PB phase introduced by the polarizer on the polarization PS is given by $\gamma_P = 2\alpha$.

Further insight into the PB phase can be obtained by interpreting it on the PS in terms of the azimuthal coordinate $\phi$. Specifically, when polarization modes located at antipodal points on the sphere traverse equal distances along geodesic arc to reach the same final state, they acquire geometric phases of equal magnitude but opposite sign. For example, a polarizer with its transmission axis oriented at $\pi/4$ radians transfers the polarization states from the north (*N*) and south (*S*) poles of the PS to a point $B$ on the sphere, as shown in Fig. 1($b$). If the azimuthal coordinate of point $B$ is $\phi_B$, then the transformation of the LCP $|LC\rangle \rightarrow |L(\pi/4)\rangle$ accumulates a geometric phase $\gamma_{NB} = -\phi_B/2$, while the transformation of the RCP $|RC\rangle \rightarrow |L(\pi/4)\rangle$ acquires a geometric phase $\gamma_{SB} = \phi_B/2$. Consequently, the net geometric phase difference generated between the two superposed modes at point $B$ is $\gamma_P = \phi_B = \pi/2$. A complete rotation of the polarizer results in two full rotations of the superposed mode along the equator of the PS, leading to a total PB phase of $\gamma_P = 4\pi$. From the above discussion, we infer that the PB phase is given by $\gamma_P = \phi = 2\alpha$. For further insight, the state transformation of the superposition state through a linear polarizer is depicted in Fig. 1($c$).

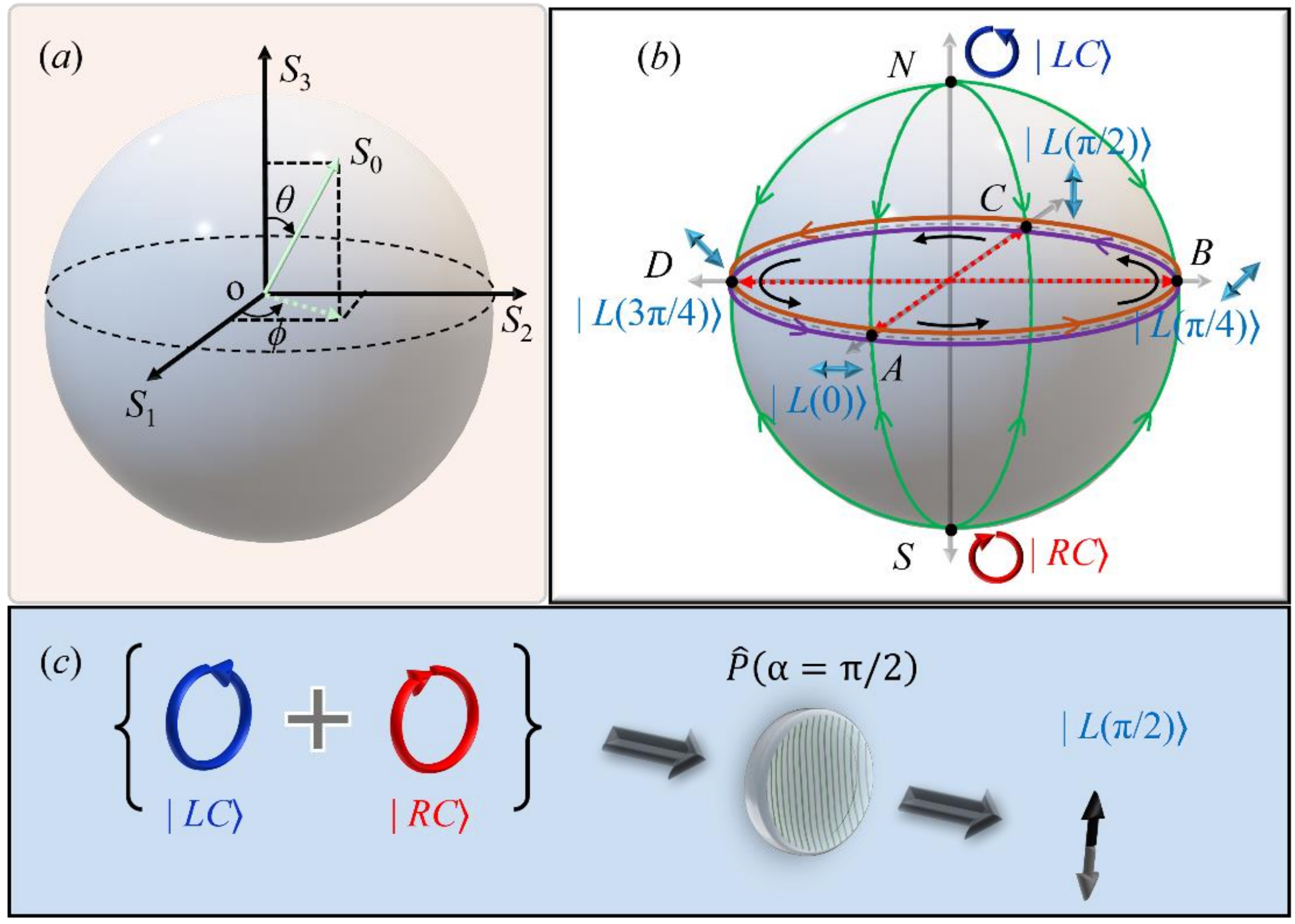

Fig. 1. Evolution of polarization on 2D parameter sphere. ($a$) Spherical coordinates and axes of Poincaré sphere. ($b$) The transportation of light from circular polarization states at poles to linear polarization states present at equator through different longitudinal directions. The brown and violet colour circular paths traversed by the linear polarization state while rotating polarizer by $2\pi$ angle. The ket-vectors: $|LC\rangle$ is left circularly polarization state, $|RC\rangle$ represents right circularly polarization state, and $|L(\alpha)\rangle$ is linear polarization state at azimuthal angle $\phi = 2\alpha$. ($d$) Experimentally light transformed from the state formed by the superposition of circular polarization states to a single linear polarization state by linear polarizer. The final state of the light is governed by the orientation of transmission axis of polarizer and acquires Pancharatnam-Berry phase (Geometric phase of light).

In 1999, M. J. Padgett and J. Courtial introduced a new 2D parameter sphere, referred to as the modal or orbital PS, which represents all possible transverse spatial modes of light of the same order [21]. On this sphere, Laguerre–Gaussian (LG) modes with equal and opposite orbital angular momenta (OAMs) are located at the poles, while Hermite–Gaussian (HG) modes lie along the equator [22]. Any state vector on the modal PS corresponds to a pure state formed by the superposition of two orthogonal structured modes within the same polarization state. For example, the state vector on the modal PS in terms of first-order optical vortices ($\ell$=±1) is given by

$$|M(\theta,\phi)\rangle = \left\{\cos\left(\frac{\theta}{2}\right)\cdot e^{-\frac{i\phi}{2}}\cdot h_1(r,\phi,z)\cdot e^{-i\phi} + \sin\left(\frac{\theta}{2}\right)\cdot e^{\frac{i\phi}{2}}\cdot h_2(r,\phi,z)\cdot e^{i\phi}\right\}. \quad (4)$$

Here, the wave function $h_i(r,\phi,z)$ describes the spatial distribution of the optical vortex nested by the laser beam. Equation 4 represents the first-order modal PS, which functions analogously to the polarization PS (Eq. 1). Higher-order modal spheres can also be constructed by considering $\ell > \pm 1$ [23]. All spatial modes on the modal PS possess non-uniform phase distributions but share a single polarization state. A modal sphere constructed from the superposition of first-order LG modes with $\ell = \pm 1$ and $p = 0$ is shown in Fig. 2($a$). State transformations on the modal PS can be implemented using an astigmatic mode converter (AMC), which imparts a geometric phase to the state on the modal PS, given by $\gamma_M = -\ell\Omega/2$ [24,25].

Another parameter sphere, referred to as the higher-order or hybrid PS, has been developed to represent all vector modes—including cylindrical vector modes, modes, and Poincaré modes—through suitable combinations of spatially structured modes and polarization states [26,27]. This sphere is constructed by combining modes from the polarization PS and the modal PS, forming a composite sphere derived from two distinct 2D Hilbert spaces. Consequently, the states on this parameter sphere are mixed states, each of which can be selectively accessed using appropriate optical gadgets. While all states on the sphere may share the same or different transverse amplitudes, they differ in both polarization and transverse phase. The generalized state on the vector PS can be mathematically expressed as a superposition of two orthogonal spatial modes coupled to orthogonal circular polarization states, as

$$|V(\theta,\phi)\rangle = \left\{\cos\left(\frac{\theta}{2}\right)\cdot e^{-\frac{i\phi}{2}}\cdot h_1(r,\phi,z)\cdot e^{i\phi\ell_1}\cdot|LC\rangle + \sin\left(\frac{\theta}{2}\right)\cdot e^{\frac{i\phi}{2}}\cdot h_2(r,\phi,z)\cdot e^{i\phi\ell_2}\cdot|RC\rangle\right\}. \quad (5)$$

In the construction of vector modes, both the type and order are determined by the transverse phase and polarization states. For example, in Fig. 2(*b*), we illustrate a vector PS whose states correspond to first-order cylindrical vector modes, constructed from two first-order LG modes with $\ell = \pm 1$ in orthogonal circular polarization states.

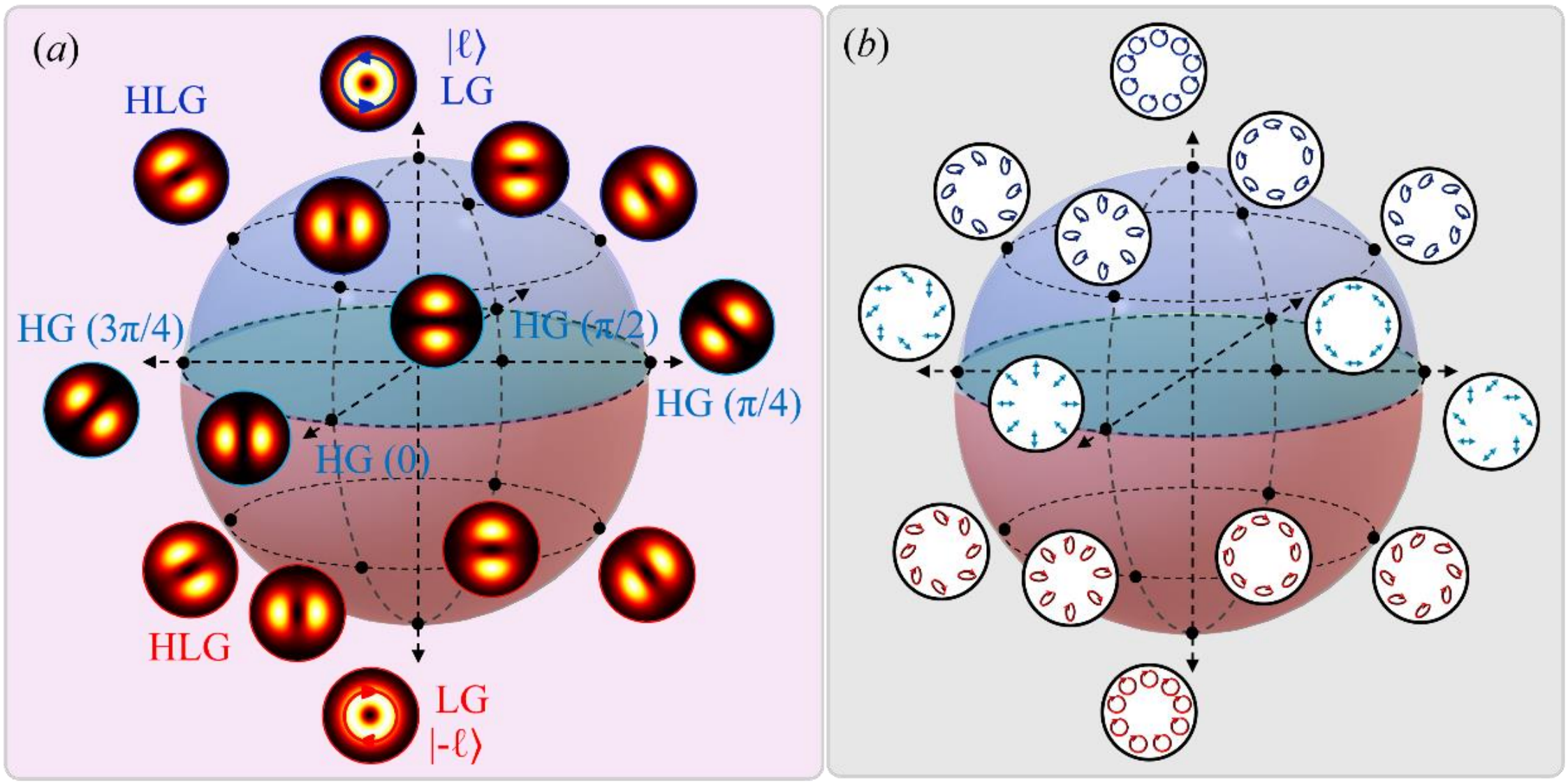

Fig. 2. (*a*) Spatial modes on the first order modal/orbital Poincaré sphere. Here, LG is Laguerre-Gaussian, HG is Hermite-Gaussian, and HLG is Hermite-Laguerre Gaussian. The arrow on the LG modes represents direction of helicity. (*b*) First order cylindrical vector modes on the higher-order Poincaré sphere.

On the higher-order PS, the geometric phase arises from two contributions: the spin angular momentum (SAM) and the orbital angular momentum (OAM) of light. This geometric phase can be expressed mathematically as $\gamma_V = -(\ell + s)\Omega/2$, and it was experimentally measured by G. Milione et al. through the transformation of vector modes on the higher-order PS using two spin–orbit converters, which are composed of an astigmatic mode converter (AMC) and a half-wave plate [28].

It is important to note that the polarization state of a superposition can be transformed on the polarization PS using suitable polarization optical gadgets without affecting the spatial modes. Similarly, spatial modes can be transformed on the modal PS using appropriate modal optical gadgets without altering the polarization states. In other words, a geometric phase can be generated between superposed modes either by changing the polarization state while keeping the spatial mode fixed, or vice versa. In the characterization of vector beams, polarization optical gadgets such as polarizers, half-wave plates, and quarter-wave plates are used to project the superposition state onto a single polarization state. For example, consider the superposition mode provided in Eq. 5 with modes have equal amplitude, i.e.,

$$|V(\pi/2,\phi) = \frac{1}{\sqrt{2}}\left\{e^{-\frac{i\phi_d}{2}}\cdot h_1(r,\phi,z)\cdot e^{i\phi\ell_1}|LC\rangle + e^{\frac{i\phi_d}{2}}\cdot h_2(r,\phi,z)\cdot e^{i\phi\ell_2}|RC\rangle\right\}. \quad (6)$$

By applying a state transformation using a polarizer, the orthogonal circular polarization states can be converted into a single linear polarization state located on the equator of the polarization PS. This state transformation can be mathematically expressed as

$$\hat{P}(\alpha)|V(\pi/2,\phi) = \frac{1}{2}\{h_1(r,\phi,z)\cdot e^{i\phi\ell_1} + h_2(r,\phi,z)\cdot e^{i\phi\ell_2}\cdot e^{i(\phi_d+2\alpha)}\}|L(\alpha)\rangle. \quad (7)$$

Now, the two orthogonal optical vortices projected onto a single polarization state $|\,L(\alpha)\rangle$ using a polarizer, and it produce a characteristic petal structure due to interference. The azimuthal orientation of the petals is determined by both the dynamical phase $\phi_d$ and the geometric phase $\gamma_P = 2\alpha$. For a given order of the vector beam, the dynamical phase defines the type of vector beam, while the petal structure rotates at a rate corresponding to the polarizer angle due to the polarization-induced PB phase. From this, we infer that the polarizer's transmission axis projects the orthogonal polarization components of the superposition state into a single polarization state, and the observed rotation of the petal structure originates from the PB phase associated with this polarization state transformation. This analysis can be readily illustrated using a radially polarized vector beam (Fig. 3). First, the dynamical phase is set to $\phi_d = 0$ so that the orientation of the petal structure aligns with the transmission axis of the polarizer. As the polarizer is rotated, the petal structure rotates along the equator of the polarization PS at the same rate, confirming the radially polarized nature of the beam [Fig. 3(*a*)]. In this case, the input superposition state has a doughnut-shaped intensity profile due to the orthogonality of the polarization states, while the output exhibits a petal pattern because the constituent modes are projected onto a single polarization state. Both modes can be decomposed into their constituent components [Fig. 3(*b*)]. A similar analysis applies to azimuthally polarized vector beams, with the dynamical phase set to $\phi_d = \pi$.

Furthermore, our analysis has been numerically verified for first and second order cylindrical vector beams [Fig. 4]. The rotation of the petal structures due to the PB phase closely matches the rotation observed when the vector beams pass through a polarizer. In practical applications, such as material processing and particle manipulation, the vector nature of these beams manifests because the orthogonal optical vortices are projected into a single polarization state, accumulating a geometric phase during interaction with matter.

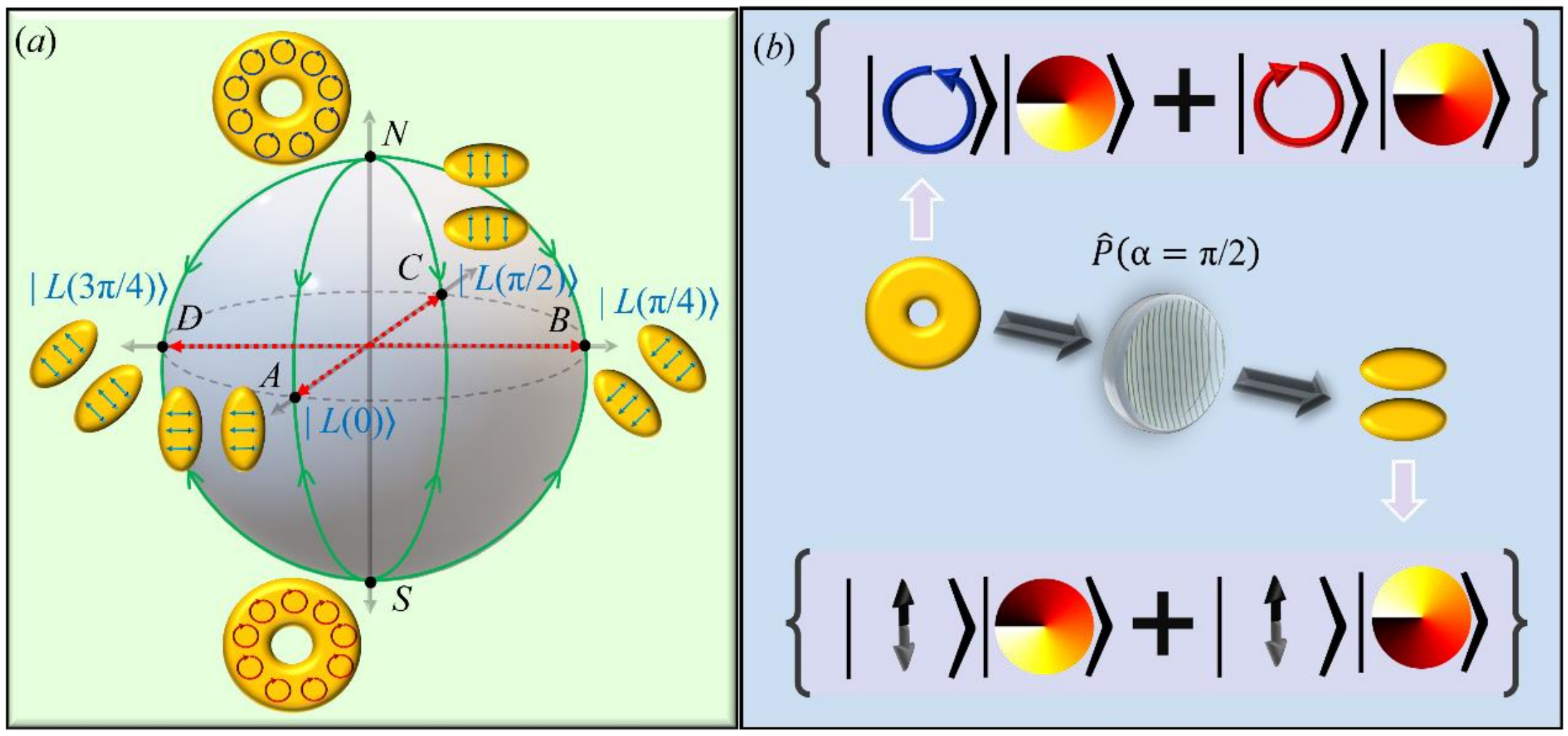

Fig. 3. Example of evolution of vector beam transportation on the polarization Poincaré sphere. (*a*) State of radially polarized first order vector beam transported on the polarization Poincaré sphere from poles to the equator at different azimuthal angles. (*b*) First order radially polarized vector beam pass through the polarizer whose transmission axis is vertical.

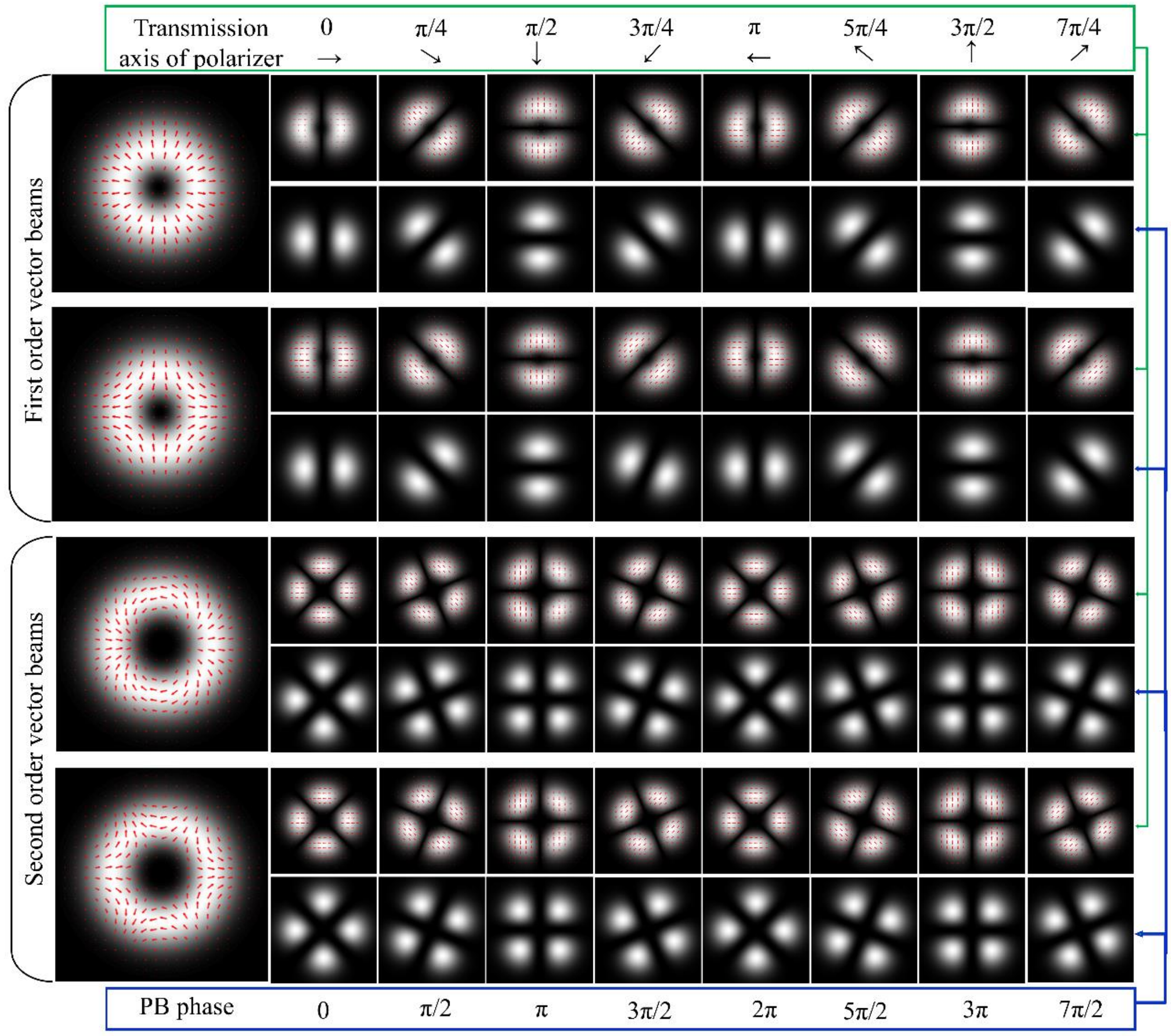

Fig. 4. Various kinds of vector beams passing through the polarizer and form a rotating petal structure. In each vector beam we have two rows; top row is the output of the polarizer when vector beam incident on it and the rotation in the petal structure created in the bottom row due to the geometric phase caused by polarizer. The red arrows represent the direction of linear polarization states. Angle and direction of polarizer given in the top and geometric phase due to corresponding angular positions of polarizer provided at the bottom of the diagram.

In conclusion, we shown that the experimentally observed vector nature of superposed orthogonal optical vortices present in the orthogonal polarization states is due to PB phase created by polarization optical gadgets. We have used polarizer as optical gadget and showed how and why it rotate the petal structure while rotating its transmission axis. While our analysis has been applied specifically to cylindrical beams and successfully verified, the same approach can be extended to Poincaré beams, other classes of vector beams, and optical quasiparticles generated using the superposition principle [13,27,29].